\documentclass[nonacm,sigconf]{acmart}
\usepackage{geometry}
\usepackage{graphicx}
\usepackage{algorithm}
\usepackage[noend]{algpseudocode}
\usepackage{amsmath}
\usepackage{lipsum}
\usepackage{placeins}
\usepackage{fancyvrb}
\usepackage{subcaption}
\usepackage{url}
\usepackage{wrapfig}
\usepackage{lscape}
\usepackage{rotating}
\usepackage{epstopdf}

\renewcommand\footnotetextcopyrightpermission[1]{}
\title{Performance modeling of a distributed file-system}
\author{Sandeep Kumar}
\authornote{Work done by author as a Master student at IISc.}
\affiliation{%
	\institution{Indian Institute of Technology Delhi}
	\city{New Delhi}
	\state{India}
}
\email{sandeep.kumar@cse.iitd.ac.in}
\begin{document}

\begin{abstract}
Data centers have become center of big data processing. Most programs running in a data center processes big data. The storage
requirements of such programs cannot be fulfilled by a single node in the data center, and hence a distributed file system is used where the the storage resource are pooled together from more than one node and presents a unified view of it to outside world. Optimum performance of these distributed file-systems given a workload is of paramount important as disk being the slowest component in the framework. Owning to this fact, many big data processing frameworks implement their own file-system to get the optimal performance by fine tuning it for their specific workloads. However, fine-tuning a file system for a particular workload results in poor performance for workloads that do not match the profile of desired workload. Hence, these file systems cannot be used for general purpose usage, where the workload characteristics shows high variation.



In this paper we model the performance of a general purpose file-system and analyse the impact of tuning the file-system on its performance. Performance of these parallel file-systems are not easy to model because the performance depends on a lot of configuration parameters, like the network, disk, under lying file system, number of servers, number of clients, parallel file-system configuration etc. We present a Multiple Linear regression model that can capture the relationship between the configuration parameters of a file system, hardware configuration, workload configuration (collectively called features) and the performance metrics. We use this to rank the features according to their importance in deciding the performance of the file-system.
\end{abstract}

\maketitle

\section{Introduction}
Due to data explosion in recent years, many computer programs often involves processing
large amount of data. For example Facebook processes about 500 PB of data daily \cite{Facebook98:online}. 
The storage capacity of a single machine is generally not enough to stored the complete data. Machines in a data centre pool their storage resources together to provide support for storing data in the range of Petabytes. Also using a single node to store large data sets creates other issues in terms of availability and reliability of data. The single machine is a single point of failure and two processes, running in parallel, working on isolated part of data cannot do it in a parallel way affecting the performance. 

\begin{figure}[ht!]
	\centering
	\includegraphics[scale=.5]{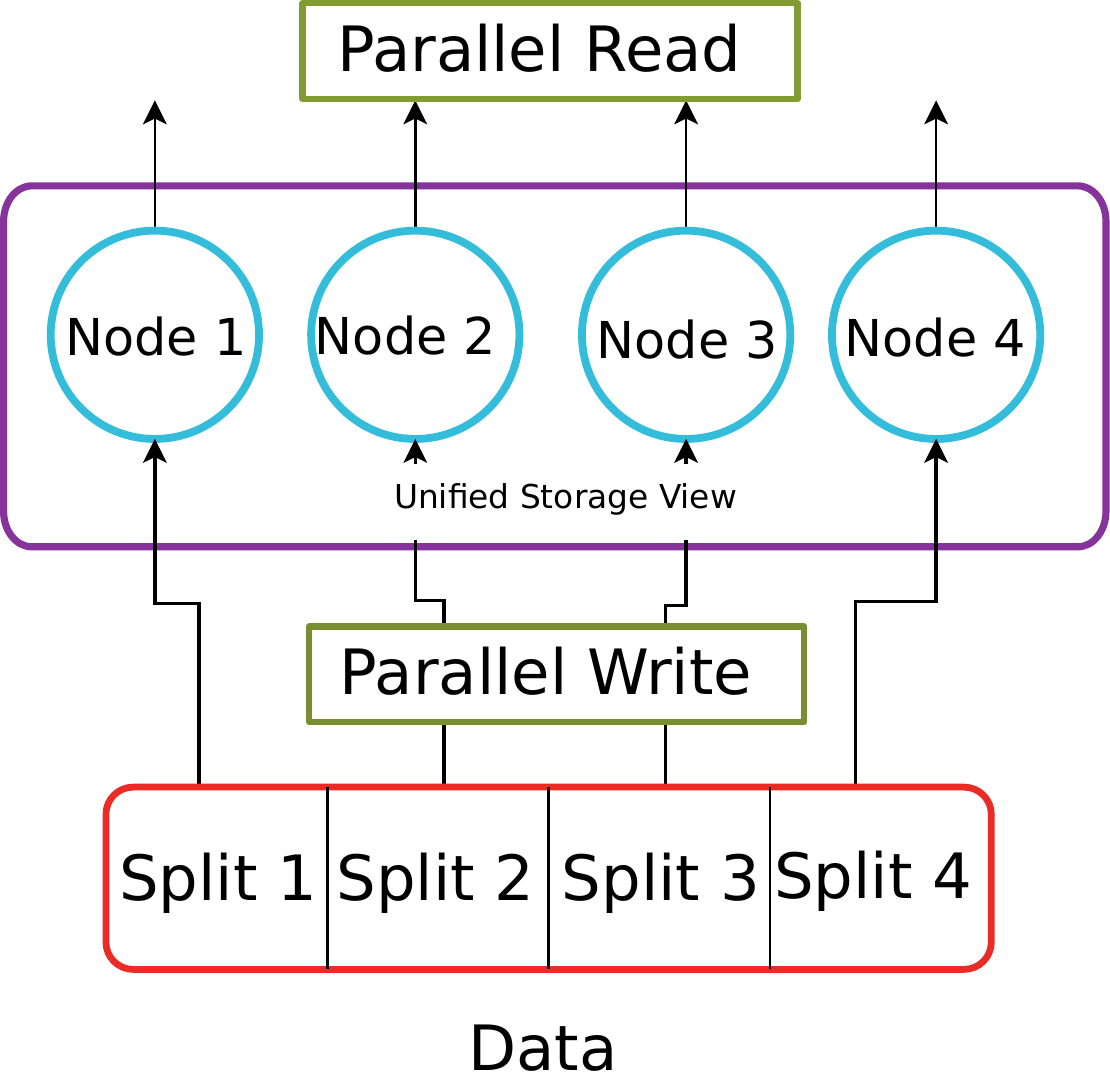}
	\caption{Typical working of a parallel file-system.}
	\label{fig:pfs_general}
\end{figure}
One way to achieve this possible is to use a parallel file system. Figure \ref{fig:pfs_general} shows a general architecture of a parallel file-system. Machines can contributed storage resources in a pool which is then used to create a virtual disk volume. The file being stored on the file-system is stripped and stored part-wise on different machines. This allows for write, and later read, operations to happen in parallel; boosting the performance significantly. 
Most of the parallel file system can do auto load balancing of the data on the servers (nodes which are used in the parallel file system), which makes sure that no single node will become a bottleneck. Using replication (where there are a number of copies of a file stored on different servers), we can achieve high availability of the data. Hadoop Distributed File System (HDFS), LustreFS, GlusterFS \cite{GlusterS15:online,LustreWi86:online,LustreWi39:online,5496972} are some of
the common parallel file system. All of them have the same purpose but differ significantly in their architecture. Some of them are used for general purpose storage (Lustre Gluster),
whereas some are optimized for very specific kind of workloads (HDFS).

\subsection{Motivation}
Using a parallel file system effectively requires an understanding of the architecture and the
working of the file system plus a knowledge about the expected workload for the file system. A
major problem with any given parallel file system is that they come up with many configurable
options which requires an in depth knowledge about the design and the working of the file system
and their relationship with the performance.
Together with the hardware configuration, they overwhelm the user with the number of
ways one can configure the cluster, consequently, an user generally ends up using the
filesystem with the default options and this may result in an under-utilization of the resources
in the cluster and a below average performance. After experiencing low performance, user
tries to upgrade the system without really knowing the cause of the performance bottleneck,
and many times ends up making a wrong decision, like buying more costly hard drives for the
nodes when the real cause of the performance bottleneck is the network.

Through our model we have identified some of the key configuration parameters in the
Gluster File system, hardware configuration and the workload configuration (features) that
significantly affect the performance of the parallel file system and, after analysing their relationship with each other,
 we were able to rank the features according to their importance in
deciding the performance. We were also able to build a prediction model which can be used to
determine the performance of the file system given a particular scenario, i.e. when we know
the hardware and file system configuration.
\subsection{Contribution}
Given the importance of each of the features in a cluster, we can weigh the positive and negative
points of a given design for such a cluster system and choose a design that satisfy our
requirements.
The result obtained from the analysis can be used to analyse some of the design options available and select the best possible among them, given some initial requirements. Based on the requirement, such as, the cluster should be dense and low power; i.e. it should take less
space on a rack and use less power than a traditional cluster system; the network connectivity option was limited to Gigabit (InfiniBand connectors take a lot of space, thus violating the condition of the dense systems). To make the system use less power, we have to choose a suitable CPU for the system, which uses less power than the conventional CPU, and is not a bottleneck in the system. The list of possible CPUs are shown in the table \ref{tab:cpu_perf}.

\begin{table}[ht!]
	\begin{tabular}{c|c|c}
		\textbf{CPU}    & \textbf{Points}  & \textbf{Power Usage} \\ \hline
		Low power Atom & 916 & 8.5 Watt\\
		Low power Xeon E3-1265	 & 8306 & 45 Watt\\
		Low power Xeon E3-1280& 9909 & 95 Watt\\
		Xeon E5430& 3975 & 80 Watt\\
		Xeon E5650 & 7480 & 95 Watt\\	
	\end{tabular}
	\caption{ Choice of the CPU along with their processing power (mentioned as points as
		calculated by the PassMark Benchmark) and Power Usage. \protect \cite{PassMark69:online}}
	\label{tab:cpu_perf}
\end{table}

As seen from the result in table \ref{tab:perf_ocean}, we can see that InfiniBand is big plus point for the performance of a distributed program and the processing power of the CPU is the actual bottleneck. However if we move to a Gigabit connection and towards a denser server cluster (12 nodes or 24 nodes in a 3U rack space), then the network performance is the limiting factor and CPUs
with low processing power are well capable of handling the load. Atom CPU have the lowest power consumption, but their processing power is also very low and even when gigabit is used the CPU will be the bottleneck, so it was not a suitable option.

After weighing the processing capability and the power usage of all the CPUs, low power Xeon
E3-1265 Servers with Gigabit connectivity has been chosen. Why?

Add Section introductions here.

\section{Background}
There are a number of parallel file systems available for use in a cluster. Some of them can be
used for general purpose storage whereas some of them are optimized for some specific type
of usage.
While selecting the file system for our experiment, along with the general requirements of a
file system like consistency and reliability, we also focused on the following:

\begin{itemize}
\item Usage scenarios: We checked in which scenarios this file system can be used and
whether it is adaptable to the different workloads demands of the user or if it is designed
for only one type of specific workload.
\item Ease of installation: It should be easy to install, like what information it requires during
the installation and whether a normal user will have access to it.
\item Ease of management: How easy it is to manage the file system, so that it can be managed
without special training
\item Ease of configuration: A file system will require many tweaks during its life time. So
it must be very easy to configure the file system and also if it can be done online (when
the virtual volume is in use) then that is a huge plus point.

\item Ease of monitoring: An admin monitoring the usage of the file system should me able
to make better decisions about load balancing and whether the cluster needs a hardware
upgrade.
\item Features like redundancy, striping etc.: The parallel file systems are generally installed
on commodity hardware, and as the size of cluster grows chances of a node failing also
increases. So the file system must support feature like replication, which make sure that
even if some of the nodes fails there is less chance of data corruption. Striping helps in
reading a file in parallel which results in a huge performance gain.
\item Point of failures and bottlenecks: We looked at the architecture of the file system to
figure out how many points of failure the file system has or what can result in a bottleneck
for the performance.
\end{itemize}

We studied the features of some of the most common filesystem being used as of today.
The first was the Lustre File system, which is used in some of the worlds fastest Super
computers \cite{LustreWi86:online,LustreWi39:online}. Improvement in the disk operation is obtained by separating the metadata
from the data. To handle the metadata there is a separate metadata server used, whereas data
is stored in separate servers. This metadata server is used only during resolving a file path and
in the permission checks. During the actual data transmission, client directly contacts the data
storage server.
Using a single metadata server increases the performance but it also becomes a performance
bottleneck as all the requests for any file operation has to go through this. This is also a single
point of failure, because if the metadata server is down, technically whole file system is down
(however this can be tackled by configuring a machine as the backup of the metadata server
which kicks in when the main metadata server fails).
Installing the Luster file system is not an easy process as it involves applying several patches
to the kernel.

The next was the Hadoop distributed file system (HDFS) which is generally used along
with the Hadoop, an open source framework for distributed computing. Hadoop is generally
used to process large amount of data, so HDFS is optimized to read and write large files in a
sequential manner. It relaxes some of the POSIX specifications to improve the performance.
The architecture is very similar to that of the Lustre file system. It also separates the metadata from the data and a single machine is configured as the metadata server, and therefore it suffers
from the same problem of performance bottleneck and single point of failure.
Lastly we looked into Gluster File system, which is a evolving distributed file system with
a design that eliminates the need of a single metadata server. Gluster can be easily installed on
a cluster and can be configured according to the expected workload. Because of these reasons
we choose Gluster file system for further analysis.

\subsection{Gluster File-System}
\begin{figure}
	\centering
	\includegraphics[scale=.5]{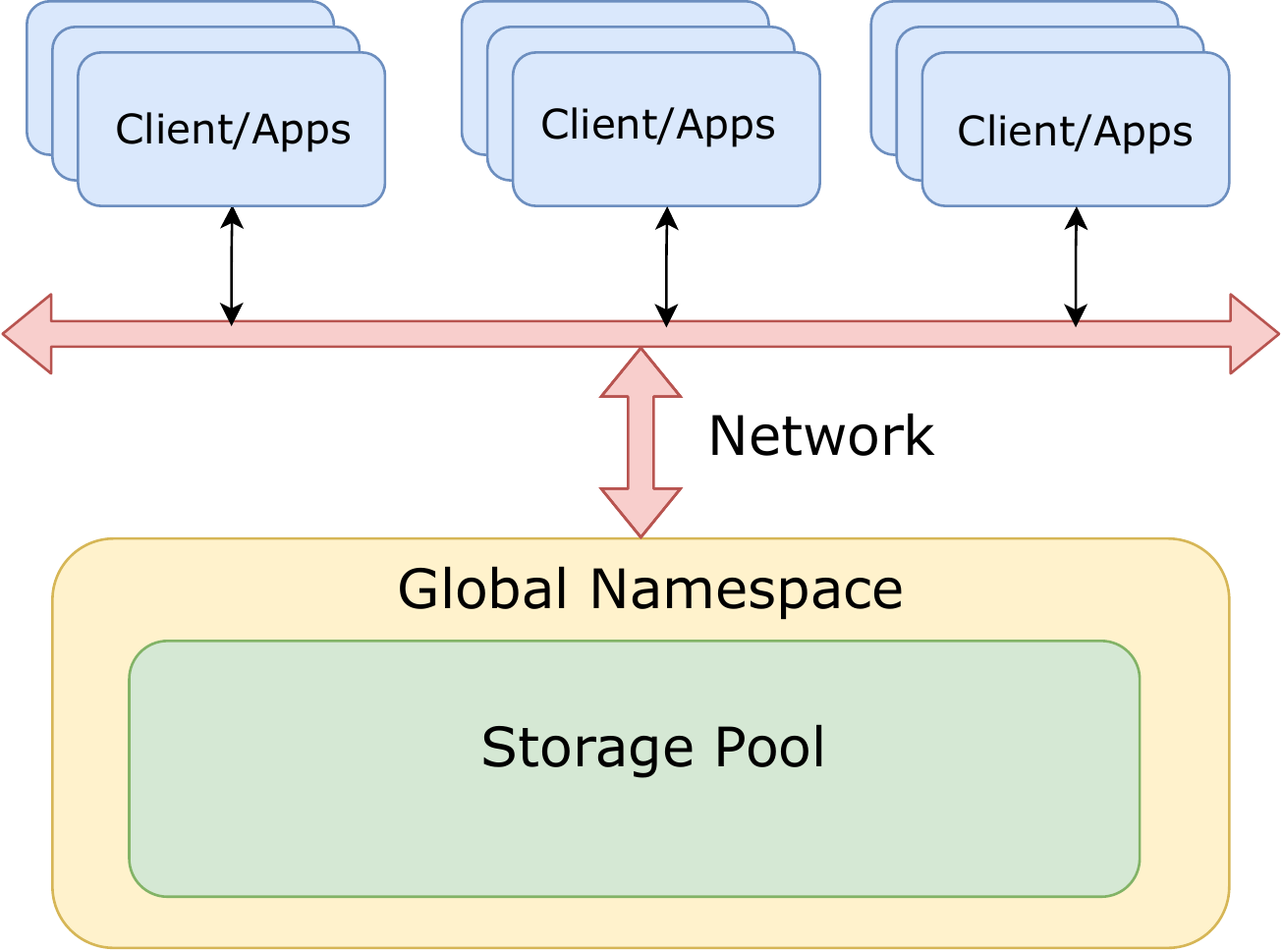}
	\caption{Architecture of the Gluster File system \protect \cite{GlusterS15:online}}
	\label{fig:result_4}
\end{figure}

Gluster file system is a distributed file system which is capable of scaling up to petabytes of
storage and provide high performance disk access to various type of workloads.
It is designed to run on commodity hardware and ensure data integrity by replication of data
(user can choose to opt out of it). The Gluster file system differs from the all of the previous file systems in its architecture
design.
All of the previous file systems increased the I/O performance by separating out the data
from the metadata by creating a dedicated node for handling the metadata.
This achieved the purpose but during high loads, this single metadata server can be reason
for a performance bottleneck and is also a single point of failure.
But in Gluster file system there is no such single point of failure as the metadata is handled by
the under lying file system on top of which the Gluster file system is installed, like ext3, xfs,
ext4 etc. The main components of the Gluster architecture is:
\begin{itemize}
\item Gluster Server: Gluster server needs to be installed on all the nodes which are going to
take part in creation of a parallel file system.
\item Bricks: Bricks are the location on the Gluster servers which take part in the storage
pool, i.e. these are the location that is used to store data.
\item Translators: Translators are a stackable set of options for virtual volume, like Quick
Read, Write Behind etc.
\item Clients: Clients are the node that can access the virtual volume and use it for storage
purpose. This virtual volume can be mounted using a number of ways, the most popular are, Gluster
Native Client, NFS, CIFS, HTTP, FTP etc.
\end{itemize}
Gluster has an inbuilt mechanism for file replication and striping. It can use Gigabit as well
as InfiniBand for communication or data transfer among the servers or with the clients.
It also gives options to profile a volume through which we can analyse the load on the volume.
Newer version of Gluster supports Geo Replication, which replicates the data in one cluster to an another cluster located far away geographically.
Because of these features, we chose to do further analysis on the Gluster File system.

\section{Experiments}
Performance of a parallel file system depends on the underlying hardware, the file system
configuration and the type of the work load for which it is used.
\subsection{Feature Selection}
Hardware configurable features are shown in Table \ref{tab:hw_config_param}.
\begin{table}[ht!]
	\begin{tabular}{c|c}
		\textbf{Parameter}    & \textbf{Description} \\ \hline
		Network Raw Bandwidth & Network Speed                     \\
		Disk Read Speed       & Sequential and Random                     \\
		Disk Write Speed      & Sequential and Random                     \\
		Number of Servers     &        File-System servers              \\
		Number of clients     &                     Clients accessing Servers\\
		Replication & No. of copies of a file block\\
		Striping & No. of blocks in which a file is split\\
	\end{tabular}
\caption{Hardware and Software configurable Parameters}
\label{tab:hw_config_param}
\end{table}

GlusterFS configurable parameters:
\begin{enumerate}
\item Striping Factor: Striping Factor for a particular Gluster volume. This determines the number of parts in which a file is divided.
\item Replication Factor: Replication Factor for a particular Gluster volume. This determines the number of copies for a given part of a file.
\item Performance Cache: Size of the read cache.
\item Performance Quick Read: This is a Gluster translator that is used to improve the performance of read on small files.
On a POSIX interface, OPEN, READ and CLOSE commands are issued to read a file.
On a network file system the round trip overhead cost of these calls can be significant.
Quick-read uses the Gluster internal get interface to implement POSIX abstraction of
using open/read/close for reading of files there by reducing the number of calls over
network from n to 1 where n = no of read calls + 1 (open) + 1 (close).
\item Performance Read Ahead: It is a translator that does the file prefetching upon reading
of that file.
\item Performance Write Behind: In general, write operations are slower than read. The write-behind translator improves
write performance significantly over read by using ”aggregated background write” technique.
\item Performance io-cache: This translator is used to enable the cache. In the client side it
can be used to store files that it is accessing frequently. On the server client it can be
used to stores the files accessed frequently by the clients.
\item Performance md-cache: It is a translator that caches metadata like stats and certain extended attributes of files
\end{enumerate}
Workload features:
\begin{itemize}
\item Workload Size:
Size of the total file which is to be read or written from the disk (virtual volume in case
of Gluster).
\item Block Size:
How much data to be read and written in single read or write operation
\end{itemize}

\subsection{Performance Metrics}
The performance metrics for the Gluster file system under observation are:
\begin{itemize}
\item Maximum write Speed
\item Maximum Read Speed
\item Maximum Random Read Speed
\item Maximum Random Write Speed
\end{itemize}
These performance metrics were chosen because they captures most of the requirements
of the application.

\subsection{Experiment Setup}
We have used various tools to measure the different aspects of the file systems and the cluster,
whose results were later use to generate the file system model.
The tools used were as follow:

\begin{itemize}
\item Iozone: Iozone is a open source file system benchmark tool that can be used to produce and measure a
variety of file operation.
It is capable of measuring various kind of performance metric for a given file system like:
Read, write, re-read, re-write, read backwards, read strided, fread, fwrite, random read, pread,
mmap etc.
It is suitable for bench marking a parallel file system because it supports a distributed mode in
which it can spawn multiple threads on different machine and each of them can write or read
data from a given location on that particular node.
Distributed mode requires a config file which tells the node in which to spawn thread, where
Iozone is located on that particular node and the path at which Iozone should be executed.
Example of a config file (say iozone.config):
\begin{verbatim}
node1 /usr/bin/iozone /mnt/glusterfs
node2 /usr/bin/iozone /mnt/glusterfs
node3 /usr/bin/iozone /mnt/glusterfs
node4 /usr/bin/iozone /mnt/glusterfs
\end{verbatim}
This config file can be used to spawn 4 threads on node1, node2, node3 and node4 and execute
iozone at a given path (on which the parallel file system or the virtual volume is mounted)
Some of the option of the IOzone tools that was used

\begin{itemize}
\item $\pm m$: run the iozone tool in distributed mode
\item -t : No of thread to spawn in distributed mode
\item -r : Block size to be used during the bench-marking (It is not the block size of the file system)
\item -s : Total size of the data to be written to the file system.
\item -i : This option is used to select the test from test suite for which the benchmark is supposed
to run.
\item -I : This options enables the O SYNC option, which forces every read and write to come from
the disk. (This feature is still in beta and does not work properly on all the clusters)
\item -l : Lower limit on number of threads to be spawned
\item -u : Upper limit on the number of threads to be spawned
\end{itemize}
Example:
\begin{verbatim}
iozone -+m iozone.config -t <no of threads>
-r <block size> -s <workload size>
-i <test suits> -I

\end{verbatim}
It does not matter on which node this command is issued, till the iozone.config file contains
valid entries. Iozone can be used to find out the performance of the file system for small and large files by
changing the block size. Suppose the total workload size is 100 MB, then if the block size is 2 KB then its like reading
or writing 51200 files to the disk. Whereas if the block size is 2048 KB then its like reading
or writing to 50 files.

\item Iperf: Suppose the total workload size is 100 MB, then if the block size is 2 KB then its like reading
or writing 51200 files to the disk. Whereas if the block size is 2048 KB then its like reading
or writing to 50 files.

\begin{verbatim}
iperf -s
\end{verbatim}
then on the client side connect to the server by issuing the command:
\begin{verbatim}
iperf -c node1
\end{verbatim}
(assuming there is an entry for node1 in the /etc/hosts file).
\item Matlab and IBM SPSS, is used to create the model from the generated data and to calculate
the accuracy of the model.
\end{itemize}

\subsection{Cluser Setup}
\subsubsection{\textit{Fist} Cluster}
The whole of the \textit{Fist} cluster comprises of total 36 nodes in total, out of which 1 node is
the login node and is used to store all the users data. The remaining 35 nodes are used for
computing purposes.
The configuration of the 35 nodes are: CPU: 8 core Xeon processor @ 2.66 GHz
RAM: 16 GB
Connectivity: Gigabit (993 Mb/sec),
Infiniband (5.8 Gb/sec)
Disk: 3 TB 7200 RPM (5 nodes),
250 GB 7200 RPM (All of them).
Out of these 35 nodes, we reconfigured the 5 of them with 3TB hard disk, and used them
as Gluster servers for our experiments.
Rest of the nodes were chosen to act as the client.
This setting made sure that if a client wants to write or read data to the virtual volume (mounted
via NFS or Gluster Native client from the servers), the data has to go outside the machine, and
the writing of data to a local disk (which probably will increase the speed) is avoided.
\subsubsection{Atom Cluster}
In the atom cluster there are 4 nodes. The configuration of each of the atom nodes are as
follow:
CPU: Atom, 4 Core processor @ 2 GHz
RAM: 8 GB
Connectivity: Gigabit (993 Mb/sec)
Disk: 1 TB, 7200 RPM
Clients are connected to the cluster from outside via a Gigabit connection.

\subsection{Data Generation}
The data generation for the analysis was a challenge since some settings for some of the features
vastly degrades the performance of the file system.
For example, when the block size is set at 2 KB, the speed for writing 100 MB of file (to disk
and not in the cache, by turning on the O SYNC) was 1.5 MB/sec.
The largest size of the workload in our experiment is 20GB. So the time taken to write 20GB to the disk will be around 30000 seconds, i.e 8.3 Hours.
In the worst case when number of clients is 5 and replication is also 5, the total data written to
the disk will be 500 GB (20*5*5).
The time taken to write 500 GB of data will be around 8.6 days, and including time taken for
reading, random reading and random writing the data, the total time will be ,
8.6 + 3.34+3.34+8.6 = 23.88 days.

\begin{figure}
	\centering
	\includegraphics[scale=.27]{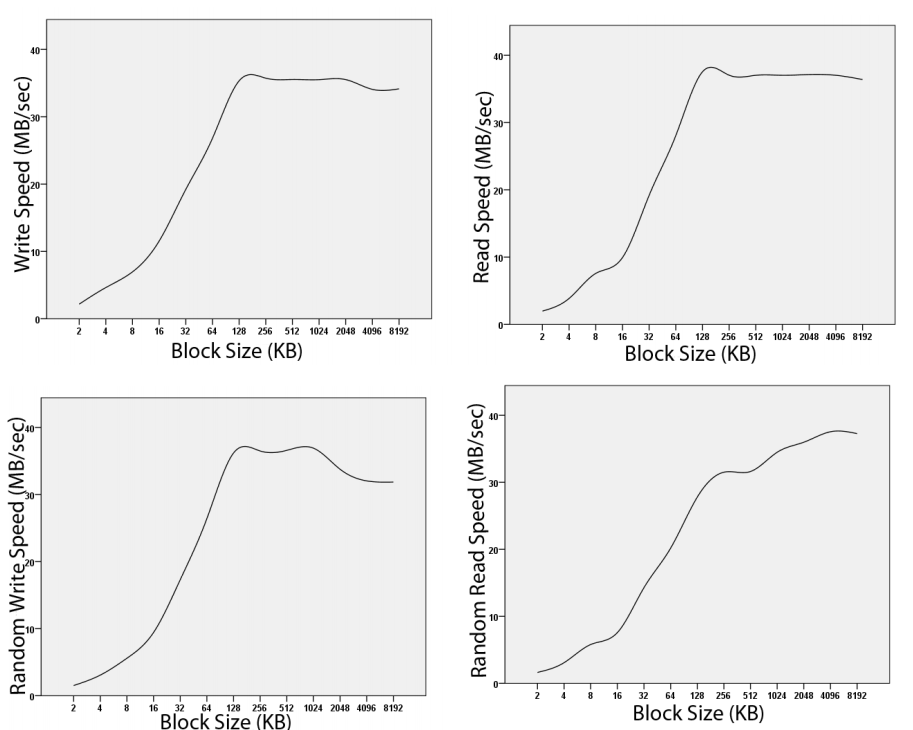}
	\caption{Relationship of the Performance metrics with the Block Size with O\_SYNC On}
	\label{fig:result_2}
\end{figure}

The read and write speed achieved above is maximum, when cache size is equal to or more
than 256 KB, as shown in Figure \ref{fig:result_2}.
If we try to model all of them together (using the normal values for the Gluster configuration),
Gluster configuration is neglected as they have negligible effect on the performance
when compared to Hardware and Workload configuration.
But they become important once the hardware and type of workload is fixed, Gluster configuration
is used to configure the File system in the best way possible.
For this reason we decided to create two models, one for the hardware and the workload configuration
and another for the Gluster configuration.

For the first model the parameters varied are listed in Table \ref{tab:FIST_cluster_variations}. The size of the workload
was dependent on the system RAM, which is 16 GB. If we try to write a file whose size is
smaller than the 16 GB then the file can be stored in the RAM and we will get a high value for
read and speed. To avoid this workload size of 20 GB was chosen.
But cache plays an important role when we try to read or write small files. We capture that
behavior of the file system by making the workload size less than the RAM.
For the second model, the workload size was fixed at 100 MB and Table \ref{tab:gluster_cluster_variations} contains the
list of the parameters and the values that they take during benchmark process.
Disk was unmounted and mounted before each test to avoid the caching effect.

\subsection{Parameter Variation}
\begin{table}[ht!]
	\begin{tabular}{c|c}
		\textbf{Parameter}    & \textbf{Values} \\ \hline
		Network 				& Gigabit and Infiniband                   \\
		Disk Read Speed       &    117 MB/sec, 81 MB/sec                  \\
		Disk Write Speed      &            148 MB/sec, 100 MB/sec          \\
		Base File System      &             Ext3, XFS         \\
		Number of Servers     &        1-.1              \\
		Number of clients     &        1-5\\
		Striping	     &                1-5 \\
		Replication     &                1-5  \\		
		Workload Size     &               100 MB, 200 MB, 500 MB,\\
		& 700 MB, 1 GB, 10 GB, 20 GB \\
	\end{tabular}
	\caption{Hardware configurable Parameters varied on \textit{Fist} Cluster}
	\label{tab:FIST_cluster_variations}
\end{table}

\begin{table}[ht!]
	\begin{tabular}{c|c}
		\textbf{Feature}    & \textbf{Values} \\ \hline
		Block Size 				& 2 KB, 4 KB, 8 KB to 8192 KB                  \\
		Performance Cache Size      &    2 MB, 4 MB, 8 MB to 256 MB                 \\
		Write Behind Translator      &           On/Off      \\
		Read Ahead Translator      &             On/Off        \\
		IO Cache Translator     &        On/Off              \\
		MD Cache Translator     &        On/Off\\
	\end{tabular}
	\caption{Gluster Parameters varied on \textit{Fist} Cluster and Atom Cluster}
	\label{tab:gluster_cluster_variations}
\end{table}

\section{Analysis}

\begin{figure}
	\centering
	\includegraphics[scale=.33]{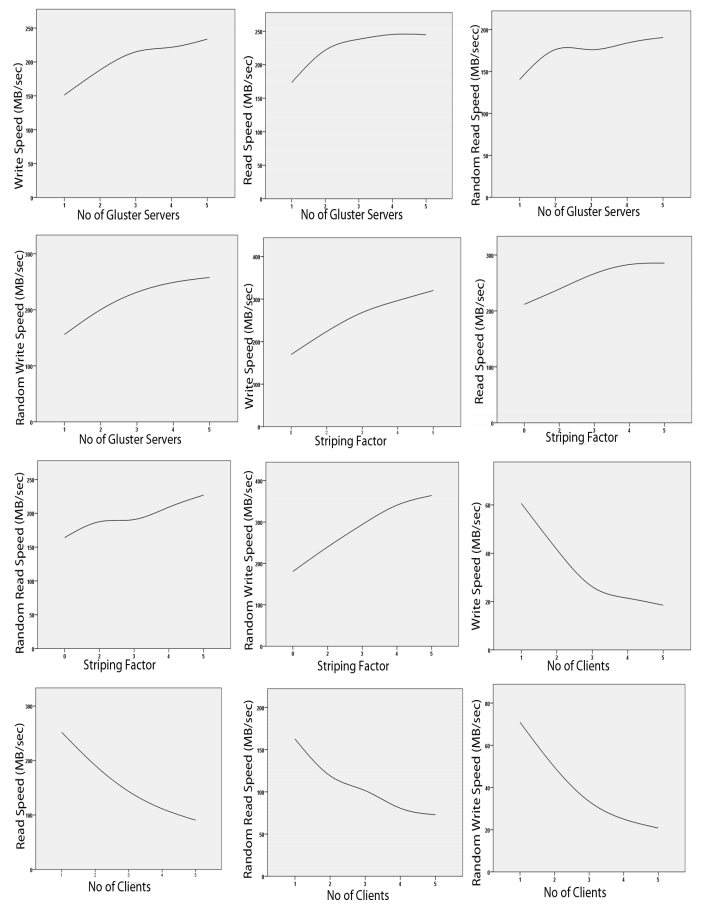}
	\caption{Figure showing an approx. linear relationship between Number of Gluster Servers,
		Striping and Number of client with the performance metrics Write Speed, Read Speed,
		Random Read Speed, Random Write Speed}
	\label{fig:result_3}
\end{figure}
As stated earlier our goal is to figure out the relationship and impact of the hardware, Gluster
and workload features on the performance features.
Multiple linear regression fits our criteria well as it can be used for:
\begin{itemize}
\item Prediction: Predicting the value of Y (the performance metric), given the cluster environment
and the Gluster configuration (X).
\item  Description: It can also be used to study the relationship between the X and Y, and in
our case we can study the impact of the various configurations on the performance, for
example, which of them affects the performance the most and which has the least effect
on the performance
\end{itemize}
\subsection{Assumptions}
Before applying multiple linear regression, the condition of linearity must be satisfied i.e. there
should be a linear relationship among the Dependent variable and the Independent variables to
get a good fit.
The relationship among the configuration features and the performance features can be
seen in Figure \ref{fig:result_3}.
\subsection{Multiple Linear Regression}
To study the relationship between more than one independent variable (features) and one dependent
variable (performance) multiple regression technique is used.
The output of the analysis will be coefficients for the independent variables, which is used to
write equation like:
\begin{equation}
\begin{aligned}
y_i = \beta_0+\beta_1x_{i1}+\beta_2x_{i2}+...+\beta_px_{ip}+\epsilon_i,i=1,..,n\\
y_i=x_i^T\beta+\epsilon_i
\end{aligned}
\end{equation}
where,\\
$y_i: $ Dependent variable’s $i^{th}$ sample (Performance metric).\\
$x_i: $ Independent variable's $i^{th}$ sample (Configuration parameter).\\
$\epsilon: $ error term\\
$\beta: $ Coefficients (output of analysis)\\
$n: $ No of data sample we have\\
$p: $ No of independent variable we have\\
In vector form:\\
\begin{math}
Y = 
\begin{pmatrix}
y_1 \\
y_2 \\
\vdots\\
y_n 
\end{pmatrix},
X = 
\begin{pmatrix}
1 & X_1^T\\
1 & X_2^T\\
\vdots & \\
1 & X_n^T
\end{pmatrix}
 = 
\begin{pmatrix}
1 & x_{11} & \hdots & x_{1p} \\
1 & x_{21} & \hdots & x_{2p} \\
\vdots& & \ddots &\vdots\\
1 & x_{n1} & \hdots & x_{np} \\
\end{pmatrix}\\
\beta = 
\begin{pmatrix}
y_1 \\
y_2 \\
\vdots\\
y_n 
\end{pmatrix}
\epsilon = 
\begin{pmatrix}
y_1 \\
y_2 \\
\vdots\\
y_n 
\end{pmatrix}
\end{math}\\
$\beta $ can be calculated by the formula:
\begin{equation}
\beta=(X^TX)^{-1}X^TY
\end{equation}

\subsection{Evaluating the Model}
The accuracy of the model can be checked by using:
\subsubsection{$R^2$:} 
It represents the proportion of the variance in dependent variable that can be explained
by the independent variables.
So, for a good model we want this value to be as high as possible. $R^2$ can be calculated in following way:
\begin{equation}
\begin{aligned}
TSS = \sum_{i=1}^{n}(y_i-\bar{y})^2, 
SSE = \sum_{i=1}^{n}(y_i-\hat{y})^2,
SSR = \sum_{i=1}^{n}(\hat{y_i}-\bar{y})^2
\end{aligned}
\end{equation}
where,
\begin{itemize}
\item TSS: Total Sum of square
\item SSE: Sum of Squares Explained
\item SSR: Sum of Square Residual
\item $y$: Dependent Variable
\item $\bar{y}$: Mean of y
\item $\hat{y}$: Predicted value for y
\item $n$: Total no of data samples available
\end{itemize}
\begin{equation}
\begin{aligned}
R^2 = 1-\frac{SSE}{TSS} = \frac{SSR}{TSS}
\end{aligned}
\end{equation}
Value of $R^2$ varies from [0,1].

\subsubsection{Adjusted $R^2$}
R2
is generally positive biased estimate of the proportion of the variance in the dependent
variable accounted for by the independent variable, as it is based on the data itself.
Adjusted R2
corrects this by giving a lower value than the R2
, which is to be expected in common data.
\begin{equation}
\hat{R^2} = 1-\Bigg( \frac{n-1}{n-k-1} \Bigg) (1-R^2)
\end{equation}
where,
\begin{itemize}
\item $n: $ Total no. of samples
\item $k: $ Total no. of features or independent variables.	
\end{itemize}
\subsection{Predictor Importance}
Predictors are ranked according to the sensitivity measure which is defined as follow:
\begin{equation}
S_i=\frac{V_i}{V(Y)}-\frac{V(E(Y|X_i))}{V(Y)}
\end{equation}
where,\\
$V(Y)$ is the unconditional output variance.
 Numerator is the expectation $E$
  over $X_{-i}$,
 which is over all features, expect $X_i$ , then $V$ implies a further variance operation on it. S is the measure sensitivity of the feature i.
S is the proper measure to rank the predictors in order of importance \cite{Saltelli:2004:SAP:994090}.
Predictor importance is then calculated as normalized sensitivity:
\begin{equation}
VI_i=\frac{S_i}{\sum_{j=1}^{k}S_j}
\end{equation}
where,
$VI$ is the predictor importance. We can calculate the predictor importance from the data itself.

\subsection{Cross Validation test}
The $R^2$ test is the measure of goodness of fit on whole of the training data. Cross validation
test is used to check the accuracy of the model when it is applied to some unseen data. For this the data set is divided into two sets called training set and the validation set. The size of these set can be decided as per the requirement. We set the size of the training set size equal to 75\% of the data set and the validation set was 25\%. The distribution of data samples to these set was completely random.

\section{Results}
Two separate analyses are done, one for the hardware and the workload configuration and another
for the Gluster configuration. In the first model, the assumption of linearity holds and hence the multiple linear regression model gives an accuracy of 75\% - 80\%. However, due to the huge impact of the cache and
block size in the Gluster configuration, the assumption of linearity is no longer valid. So we
used predictor importance analysis to rank the features according to their importance.
\subsection{Model for Hardware Configuration}
\subsubsection{Maximum Write Speed}
The output of the Multiple Linear regression analysis, i.e the coefficients corresponding to
each of the configurable feature is shown in the Table \ref{tab:coeff_write} From the coefficient table we can see
that the Random write performance is most effected by the Network, followed by Replication,
No of clients , No of servers and Striping.
The signs tell us the relationship, for example, on increasing the Network bandwidth the write
performance increases, and increasing the replication factor decreases the write speed.

\begin{table*}[ht!]
	\begin{tabular}{c|c|c|c|c}

		& \multicolumn{4}{c}{\textbf{Coefficients}}\\
		\hline
		\textbf{Feature}    & \textbf{Write	} & \textbf{Read} & \textbf{Random Read} & \textbf{Random Write}	\\ \hline
		Constant 				& -120.485                  & -142.931				& -39.676		 & -114.066							\\
		Network 				& 271.936					& 	237.502				& 	162.003		 & 	230.091							\\
		Disk Read Speed       &    $\approx 0$				& 	 $\approx 0$		& 	 $\approx 0$ & 	 $\approx 0$					\\
		Disk Write Speed      &    $\approx 0$				& 	 $\approx 0$		& 	 $\approx 0$ & 	 $\approx 0$					\\
		Base File System      &      4.419					& -4.406				& -17.888		 & -2.190							\\
		Number of Servers     &        30.729   		    &     15.890			&     14.204	 &     34.241						\\
		Number of clients     &        -34.359				& 	-21.094				& 	-16.546		 & 	-39.060							\\
		Striping	     &                10.380 			& 	3.508				& 	-.961		 & 	13.538							\\
		Replication     &                -59.182 			& 	-21.058				& 	-22.799		 & 	-60.986							\\
		Workload Size     &               -0.16				& 	-0.004				& 	-0.003		 & 	-0.13							\\
	\end{tabular}
	\caption{ The coefficients for the model for Write performance of the parallel file system}
	\label{tab:coeff_write}
\end{table*}

\begin{table}[ht!]
	\begin{tabular}{c|c|c|c|c}
		& \multicolumn{4}{c}{\textbf{Coefficients}}\\
		\hline
		\textbf{Feature}    & \textbf{Write}  & \textbf{Read} & \textbf{Rnd Read}& \textbf{Rnd Write}\\ \hline
		Adjusted $R^2$ 				& 73.5\%  &80.1\% &  74.4\%   &74.5\%      \\
		Cross Validation 				& 75.4\% & 82.5\% & 75.2\%&76.9\% \\
	\end{tabular}
	\caption{Cross validation was done by training the model on
		75\% of data and then running validation on the rest of the 25\% data}
	\label{tab:cross_write}
\end{table}

\subsubsection{Maximum Read Speed}
The output of the Multiple Linear regression analysis, i.e the coefficients corresponding to
each of the configurable feature is shown in the Table \ref{tab:coeff_read}.

From the coefficient table we can see that the Read performance is most effected by the
Network followed by Replication, No of clients and No of servers.
Replication is having an negative effect on the read performance instead of no effect or some positive effect because of the replication policy of the Gluster.
If we have specified a replication factor of say n, then when a client tries to read a file, it has
to go to every server to ensure that the replicas of that file is consistent everywhere and if it is
not consistent, read the newest version and start the replication process.
Because of this overhead the read performance of the file system drops with an increase in the
replication factor.

\subsubsection{Maximum Random Read Speed}
The output of the Multiple Linear regression analysis, i.e the coefficients corresponding to
each of the configurable feature is shown in the Table \ref{tab:coeff_random_read}. From the coefficient table we can see that the Random read performance is mostly effected by the Network, then by Replication
followed by Base file system, No of clients and No of servers.


\subsubsection{Maximum Random Write Speed}
The output of the Multiple Linear regression analysis, i.e the coefficients corresponding to
each of the configurable feature is shown in the Table \ref{tab:coeff_random_write}.

From the coefficient table we can see that the Random write performance is most effected
by the Network, followed by Replication, No of clients , No of servers and Striping.

\subsection{Model for the Gluster Configuration parameters}
Data samples were generated for 100 MB of file. Since the size of the file is much smaller
than the RAM available (8 GB), then during the benchmarks, cache effect will be present. To
avoid this, we enabled the O SYNC option in the Iozone benchmark test which will ensure
that every read and write operation is done directly to and from the disk. But we also want to
see the effect of cache when we turn on some translators that are cache dependent. So we ran
every benchmark test twice, one time with the O SYNC option ON and another time with the
O SYNC option OFF.

\subsubsection{With O\_SYNC option ON}
When the O\_SYNC option is ON, i.e. every read or write operation goes to the disk, then
the dominating feature is the block size, which is to be expected, as increasing cache size and
turning on translators will not help as we are forcing every operation to come from the disk.
We are doing it for a 100 MB file, but same kind of behaviour can be seen when dealing with
large files (size greater than the RAM size).

\subsubsection{With O\_SYNC option OFF}
\begin{figure}
	\centering
	\includegraphics[scale=.23]{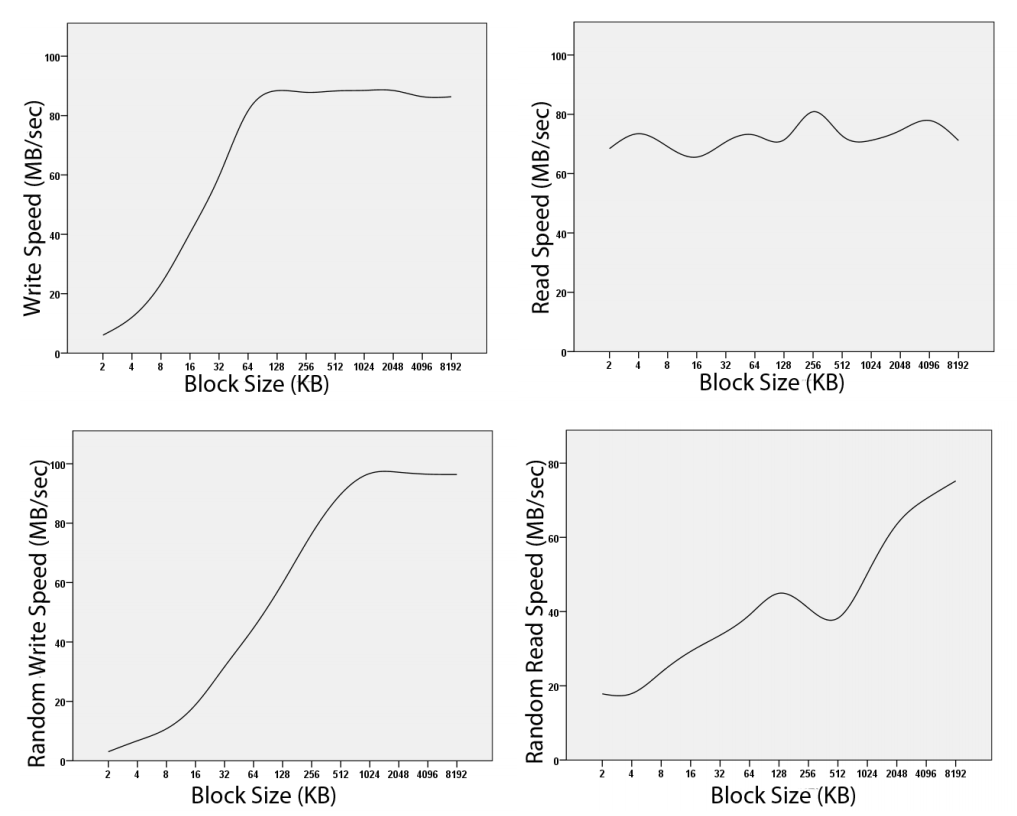}
	\caption{Relationship of the Performance metrics with the Block Size with O\_SYNC Off}
	\label{fig:result}
\end{figure}
With the O\_SYNC option off, the cache effect can be seen in the Read performance of the
system, where as the Block size is still the dominant feature
The size of the workload is 100 MB only, and still the block size is the dominating factor
but the effect of cache can be seen in the performance of read speed and random read speed
as Gluster have some translator (read ahead and quick read) that uses cache to optimize the
performance of read of the files, specially for the small files.
The performance of write and random write follows the same pattern as early.
We can see that the speed of the read operation remain the same with change in the block
size as compared to early when the O SYNC option was ON and everything has to come from
disk.

\begin{table}[ht!]
	\begin{tabular}{c|c|c|c|c}
		& \multicolumn{4}{c}{\textbf{Importance}}\\
		\hline
		& & \multicolumn{3}{c}{\textbf{O\_SYNC Off}}\\
		\hline
		\textbf{Feature}    & \textbf{Sync.} & \textbf{Rnd. Write}& \textbf{Read}& \textbf{Rnd. Read}\\ \hline
	Block Size				& $\approx 1$  & $\approx 1$    &\textbf{0.78} &\textbf{0.95}             \\
	Cache 				& $\approx 0$ & $\approx 0$ & 0.22&0.05	\\
	Rest 				& $\approx 0$ & $\approx 0$ & 0&0	\\
	\end{tabular}
	\caption{   Predictor Importance with O\_SYNC ON}
	\label{tab:imp_osync_on}
\end{table}

%
%

\section{Performance on a Real World Application}
To verify the result we compared the performance of an Ocean Modeling code in different
scenarios.
Ocean code is used in the modelling of the ocean behaviour given some initial observation. It
uses ROMS modelling tool to do the calculation and MPI to run on a cluster by spawning 48
threads. \textit{NTIMES} factor controls the number of iterations inside the code and is the major
factor in deciding the run time of the code The optimal value of NTIMES is 68400 (changes
depending upon the requirement). During its execution it processes more than 500 GB of data.
We tested its performance in situations as listed in the table \ref{tab:perf_ocean}
From the table \ref{tab:perf_ocean} we can see that the network indeed is the most dominating factor in deciding the performance of the code. On gigabit network, the performance of more number of servers
falls behind the performance of less number of servers on infiniband and as we increase the
number of servers on infiniband the time taken to complete the code decreases. We can see from the time taken to run the code that if Gigabit network is used, the performance
of the Atom CPU is very close to that of Xeon because the network was the bottleneck here.
This fact was helpful in deciding which low power servers to be bought for CASL lab.

\begin{table}[ht!]
	\begin{tabular}{c|c}
		\hline
		\textbf{Configuration}    & \textbf{Time Taken	} \\ \hline
4 Atom Servers, Gigabit					& $\approx 2311$ min, $\approx 38.5$ Hrs\\
4 Gluster Servers,  Gigabit				& $\approx 1916$ min, $\approx 31$ Hrs\\
2 Gluster Servers, Infiniband			& $\approx 478$ min, $\approx 7.9$ Hrs\\
4 Gluster Servers, Infiniband			& $\approx 282$ min, $\approx 4.7$ Hrs\\
10 Gluster Servers, Infiniband			& $\approx 144$ min, $\approx 2.4$ Hrs\\
	\end{tabular}
	\caption{Performance of the Ocean code in different scenarios}
	\label{tab:perf_ocean}
\end{table}

\section{Conclusion and Future Work}
For a first order model, Multiple Linear Regression is good enough, as the performance is
mainly decided by only one or two features (network in case of the Gluster file system). For
a much more detailed model which can explain the interaction between the various of the
features, a much more sophisticated tool is required.
The cache of the system also plays an very importance role in deciding the performance but
requires support of the file system (gluster file system optimizes the read speed using the cache
of the system).
\begin{itemize}
\item Unified Model: Because of the huge difference between the impact of the hardware configuration features
and the Gluster configuration features, a separate model was required.
A unified model can be developed by adding some bias to the Gluster configuration
features which will increase its importance and the we can incorporate all of the configuration
features in one model.
\item Automate the whole process:
Build a tool which can do necessary things like generate the data and then do appropriate
analysis on it automatically and give us the desired result.

\section{Related Work}
The machine learning algorithm used in \cite{Ganapathi:EECS-2009-181}, Kernel Canonical Correlation Analysis (KCCA) is
used to predict the performance of the file system. Apart from the features of the hardware and
file system, they also include some features from the application itself, since KCCA can tell us
the relationship between the features of the application and the features of the file system and
the hardware cluster. From some initial benchmark test we observer that only one or two of the
configuration features determine the performance of the parallel filesystem, so there is no point
in going for a very detailed model using complex methods. For the sake of completeness we are
using more features ( and not only those one which have a major impact on the performance).
A more detailed model is required to analyse the interaction between these features.

\cite{Malik2017ColocatingAC} fine tunes the performance of MapReduce in a specific scenario.
%


\bibliographystyle{ACM-Reference-Format}
\bibliography{mybib}


\begin{thebibliography}{9}


\ifx \showCODEN    \undefined \def \showCODEN     #1{\unskip}     \fi
\ifx \showDOI      \undefined \def \showDOI       #1{#1}\fi
\ifx \showISBNx    \undefined \def \showISBNx     #1{\unskip}     \fi
\ifx \showISBNxiii \undefined \def \showISBNxiii  #1{\unskip}     \fi
\ifx \showISSN     \undefined \def \showISSN      #1{\unskip}     \fi
\ifx \showLCCN     \undefined \def \showLCCN      #1{\unskip}     \fi
\ifx \shownote     \undefined \def \shownote      #1{#1}          \fi
\ifx \showarticletitle \undefined \def \showarticletitle #1{#1}   \fi
\ifx \showURL      \undefined \def \showURL       {\relax}        \fi
\providecommand\bibfield[2]{#2}
\providecommand\bibinfo[2]{#2}
\providecommand\natexlab[1]{#1}
\providecommand\showeprint[2][]{arXiv:#2}

\bibitem[\protect\citeauthoryear{??}{Fac}{[n. d.]}]%
        {Facebook98:online}
 \bibinfo{year}{[n. d.]}\natexlab{}.
\newblock \bibinfo{title}{Facebook processes more than 500 TB of data daily -
  CNET}.
\newblock
  \bibinfo{howpublished}{\url{https://www.cnet.com/news/facebook-processes-more-than-500-tb-of-data-daily/}}.
    (\bibinfo{year}{[n. d.]}).
\newblock
\newblock
\shownote{(Accessed on 12/18/2018).}


\bibitem[\protect\citeauthoryear{??}{Glu}{[n. d.]}]%
        {GlusterS15:online}
 \bibinfo{year}{[n. d.]}\natexlab{}.
\newblock \bibinfo{title}{Gluster | Storage for your Cloud}.
\newblock \bibinfo{howpublished}{\url{https://www.gluster.org/}}.
  (\bibinfo{year}{[n. d.]}).
\newblock
\newblock
\shownote{(Accessed on 12/18/2018).}


\bibitem[\protect\citeauthoryear{??}{Lus}{[n. d.]a}]%
        {LustreWi86:online}
 \bibinfo{year}{[n. d.]}\natexlab{a}.
\newblock \bibinfo{title}{Lustre - Wikipedia}.
\newblock \bibinfo{howpublished}{\url{https://en.wikipedia.org/wiki/Lustre}}.
  (\bibinfo{year}{[n. d.]}).
\newblock
\newblock
\shownote{(Accessed on 12/18/2018).}


\bibitem[\protect\citeauthoryear{??}{Lus}{[n. d.]b}]%
        {LustreWi39:online}
 \bibinfo{year}{[n. d.]}\natexlab{b}.
\newblock \bibinfo{title}{Lustre Wiki}.
\newblock \bibinfo{howpublished}{\url{http://wiki.lustre.org/Main_Page}}.
  (\bibinfo{year}{[n. d.]}).
\newblock
\newblock
\shownote{(Accessed on 12/18/2018).}


\bibitem[\protect\citeauthoryear{??}{Pas}{[n. d.]}]%
        {PassMark69:online}
 \bibinfo{year}{[n. d.]}\natexlab{}.
\newblock \bibinfo{title}{PassMark Software - CPU Benchmark Charts}.
\newblock \bibinfo{howpublished}{\url{https://www.cpubenchmark.net/}}.
  (\bibinfo{year}{[n. d.]}).
\newblock
\newblock
\shownote{(Accessed on 12/20/2018).}


\bibitem[\protect\citeauthoryear{Ganapathi}{Ganapathi}{2009}]%
        {Ganapathi:EECS-2009-181}
\bibfield{author}{\bibinfo{person}{Archana~Sulochana Ganapathi}.}
  \bibinfo{year}{2009}\natexlab{}.
\newblock \emph{\bibinfo{title}{Predicting and Optimizing System Utilization
  and Performance via Statistical Machine Learning}}.
\newblock \bibinfo{thesistype}{Ph.D. Dissertation}. \bibinfo{school}{EECS
  Department, University of California, Berkeley}.
\newblock
\urldef\tempurl%
\url{http://www2.eecs.berkeley.edu/Pubs/TechRpts/2009/EECS-2009-181.html}
\showURL{%
\tempurl}


\bibitem[\protect\citeauthoryear{Malik, Tullsen, and Homayoun}{Malik
  et~al\mbox{.}}{2017}]%
        {Malik2017ColocatingAC}
\bibfield{author}{\bibinfo{person}{Maria Malik}, \bibinfo{person}{Dean~M.
  Tullsen}, {and} \bibinfo{person}{Houman Homayoun}.}
  \bibinfo{year}{2017}\natexlab{}.
\newblock \showarticletitle{Co-locating and concurrent fine-tuning MapReduce
  applications on microservers for energy efficiency}.
\newblock \bibinfo{journal}{\emph{2017 IEEE International Symposium on Workload
  Characterization (IISWC)}} (\bibinfo{year}{2017}), \bibinfo{pages}{22--31}.
\newblock


\bibitem[\protect\citeauthoryear{Saltelli, Tarantola, Campolongo, and
  Ratto}{Saltelli et~al\mbox{.}}{2004}]%
        {Saltelli:2004:SAP:994090}
\bibfield{author}{\bibinfo{person}{Andrea Saltelli}, \bibinfo{person}{Stefano
  Tarantola}, \bibinfo{person}{Francesca Campolongo}, {and}
  \bibinfo{person}{Marco Ratto}.} \bibinfo{year}{2004}\natexlab{}.
\newblock \bibinfo{booktitle}{\emph{Sensitivity Analysis in Practice: A Guide
  to Assessing Scientific Models}}.
\newblock \bibinfo{publisher}{Halsted Press}, \bibinfo{address}{New York, NY,
  USA}.
\newblock
\showISBNx{0470870931}


\bibitem[\protect\citeauthoryear{Shvachko, Kuang, Radia, and Chansler}{Shvachko
  et~al\mbox{.}}{2010}]%
        {5496972}
\bibfield{author}{\bibinfo{person}{K. Shvachko}, \bibinfo{person}{H. Kuang},
  \bibinfo{person}{S. Radia}, {and} \bibinfo{person}{R. Chansler}.}
  \bibinfo{year}{2010}\natexlab{}.
\newblock \showarticletitle{The Hadoop Distributed File System}. In
  \bibinfo{booktitle}{\emph{2010 IEEE 26th Symposium on Mass Storage Systems
  and Technologies (MSST)}}. \bibinfo{pages}{1--10}.
\newblock
\showISSN{2160-195X}
\urldef\tempurl%
\url{https://doi.org/10.1109/MSST.2010.5496972}
\showDOI{\tempurl}


\end{thebibliography}

\end{itemize}

\end{document}